\documentclass[pre,aps,twocolumn]{revtex4}
\usepackage{epsfig}
\usepackage{bm} 
\newcommand{\B}[1]{{\bm{#1}}}
\newcommand{\C}[1]{{\mathcal{#1}}}

\newcommand{\Onecol} {\begin{widetext} \onecolumngrid} 
\newcommand{\Twocol} {\end{widetext} \twocolumngrid} 

\newcommand{\be}{\begin{equation}}
\newcommand{\ba}{\begin{array}}
\newcommand{\bea}{\begin{eqnarray}}
\newcommand{\bfi}{\begin{figure}}
\newcommand{\ee}{\end{equation}}
\newcommand{\ea}{\end{array}}
\newcommand{\eea}{\end{eqnarray}}
\newcommand{\efi}{\end{figure}}

\newcommand{\ra}{\right\rangle}
\newcommand{\la}{\left\langle}
\begin{document} 
\title{ Drag Reduction in Homogeneous Turbulence by Scale-Dependent
Effective Viscosity}
\author{Roberto Benzi$^{1,2}$, Emily S.C. Ching$^{2}$, and Itamar
Procaccia$^{2,3}$} 
\affiliation{$^1$ Dip. di Fisica and INFM, Universit\`a ``Tor
Vergata", Via della Ricerca Scientifica 1, I-00133 Roma, Italy,\\
$^2$Dept. of Physics, The Chinese University of Hong Kong, Shatin, Hong
Kong,\\ 
$^3$ Dept. of Chemical Physics, The Weizmann Institute of Science, Rehovot,
76100 Israel.} 
\begin{abstract} 
The phenomenon of drag reduction by polymer additives had been studied in
simulations 
on the basis of non-Newtonian fluid mechanical models that take into account
the 
field of polymer extension (conformation tensor) and its interaction with
the velocity field.
Drag reduction was found in both homogeneous and wall bounded turbulent
flows. In the latter case
it was shown recently that the notion of scale-dependent effective viscosity
allows quantitative
predictions of the characteristics of drag reduction in close correspondence
with experiments. 
In this paper we demonstrate that also drag reduction in homogeneous
turbulence is usefully
discussed in terms of a scale-dependent viscosity. In other words, the
essence of the phenomena under study can
be recaptured by an ``equivalent" equation of motion for the velocity field
alone, with 
a judiciously chosen scale-dependent effective viscosity that succinctly
summarizes the important aspects
of the interaction between the polymer conformation tensor field and the
velocity field. We 
will also clarify here the differences between drag reduction in homogeneous
and wall bounded 
flows. 
\vskip 0.2cm 
\end{abstract} 
\maketitle 
\section{Introduction}
\label{section1} 
\begin{figure} 
\centering 
\includegraphics[width=.5\textwidth]{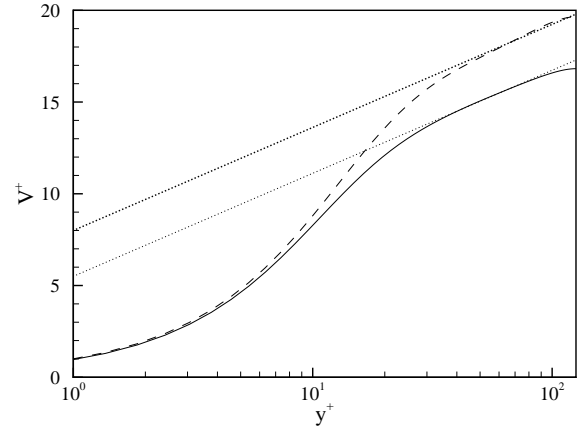}
\caption{Mean velocity profile of the FENE-P (dashed line) and of the
Navier-Stokes equations (solid line)
in wall bounded channel
flow as a function of the reduced distance from the wall. The relative
increase 
of the mean velocity (indicated by the asymptotic straight lines) is the
phenomenon of drag reduction in wall bounded flows.}
\label{fig1} 
\end{figure}

The addition of long chained polymers to turbulent flows can result in a
significant 
reduction in the drag \cite{00SW,69Lu,75Virk,90Ge}. The phenomenon had been
discovered in 1949 \cite{46Toms}
and had since attracted large amount
of attention, with much of the experimental literature reviewed and
systematized by Virk \cite{75Virk}; the amount of drag
depends on the characteristics of the polymer and its concentration, but
cannot 
exceed an asymptote known as the ``Maximum Drag Reduction" curve which is
independent 
of the polymer's concentration or its characteristics. The understanding of
this phenomenon 
had seen significant progress recently. A first step in
forming a new understanding were direct numerical simulations of model
equations of 
viscoelastic flows, both in wall bounded and in homogeneous turbulence
\cite{97THKN,98DSB,00ACP}. The Oldroyd-B
and the FENE-P models first, and then simplified models like shell models
and Burger's like 
models \cite{03BDGP,03BP,03BCHP}, all exhibited drag reduction as a result
of including the interaction between the
velocity field and a second field representing the polymer (end-to-end)
conformation 
tensor, see Figs. 1 and 2. In homogeneous turbulence drag reduction is
exhibited 
as the increase in the root-mean-square (rms) velocity fluctuations at
scales larger than 
the Lumley scale defined as the scale for which the eddy turnover time is of
the order of the polymer relaxation time.
The intermediate scale rms energy fluctuations are suppressed due to
transfer of energy to the
polymers. In wall bounded turbulence drag reduction entails an increase in
the mean velocity 
for a given pressure head, see Fig. \ref{fig1}. Here the Reynolds stress at
the intermediate 
scales is suppressed \cite{03ACLPP}; we will argue however that there is a
difference between the
increase in the rms velocity fluctuations at large scales in homogeneous
flows and the increase in mean velocity in
wall bounded flows; the former disappears when the system size goes to
infinity (for a fixed Lumley scale). In the latter
case an increase in the mean velocity near the wall
(small and intermediate scales) does not disappear with increasing the
system's size. This difference is fundamental to the different symmetries at
play,
the Galilean invariance in the case of the wall bounded flow vs.
translational
invariance in the case of homogeneous flows. Nevertheless we will argue
below that the two cases can be discussed in similar physical terms.
\begin{figure} 
\centering 
\includegraphics[width=.5\textwidth]{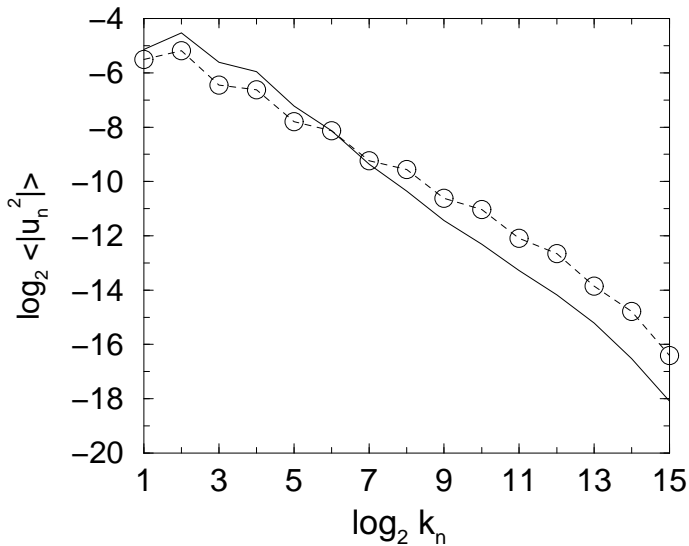}
\caption{Energy spectrum of the SabraP model (line) and the Sabra
model (dashed line with symbols) for $\nu=10^{-7}$. The relative increase of
the 
energy spectrum at small values of $n$ is the phenomenon of drag reduction
in 
homogeneous turbulence in general and in shell
models in particular, see Sect. \ref{homo} for details.}
\label{spec} 
\end{figure} 
In a recent paper it was shown that drag reduction in wall bounded flows can
be 
conveniently discussed in terms of a `scale-dependent' effective viscosity.
The aim of the present paper is to demonstrate that this notion is also
useful 
in the context of homogeneous turbulence.
In doing so we aim at simplifying the theoretical description, eliminating
the explicit 
presence of a second field in the equations of motion, leaving the velocity
field alone. 
The eliminated field, which represents the conformation tensor of the
polymers, remains only
as an effective viscosity in the equation of motion. Needless to say, this
effective viscosity
cannot be a number, since the amount of energy transferred from the velocity
field to the polymer is
strongly scale dependent; in homogeneous turbulence this transfer achieves a
maximum near the Lumley scale.
In wall bounded flows the degree of interaction between the polymers and the
velocity field is 
a strong function of the distance from the wall, and so is therefore the
effective viscosity.
Of course, in a full theory a scale-dependent scalar viscosity is not
sufficient either, due to the anisotropy of
the polymer end-to-end extension tensor. We would like to demonstrate
however that at least
in the model equations a surprising proportion of the essential physics can
be captured in 
terms of a simple notion of a scale-dependent viscosity which surrogates the
existence of the second field. This thinking goes back to some observations
a few years ago 
regarding the importance of space-dependent viscosity even in the stability
of laminar flows 
\cite{01GLP,02GLP}.
In Sect. 2 we review the two-field models in which drag reduction had been
demonstrated 
in numerical simulations. In Sect. 3 we present the
reduction to velocity-alone models with scale-dependent viscosity.
In Sect. 4 we present a discussion of the large system size limit
and underline the difference between homogeneous and wall bounded flows.
Sect. 5 is dedicated to a short summary and conclusions.
\section{Shell model for drag reduction in homogeneous turbulence}
\label{section2} 
\begin{figure} 
\centering 
\includegraphics[width=.5\textwidth]{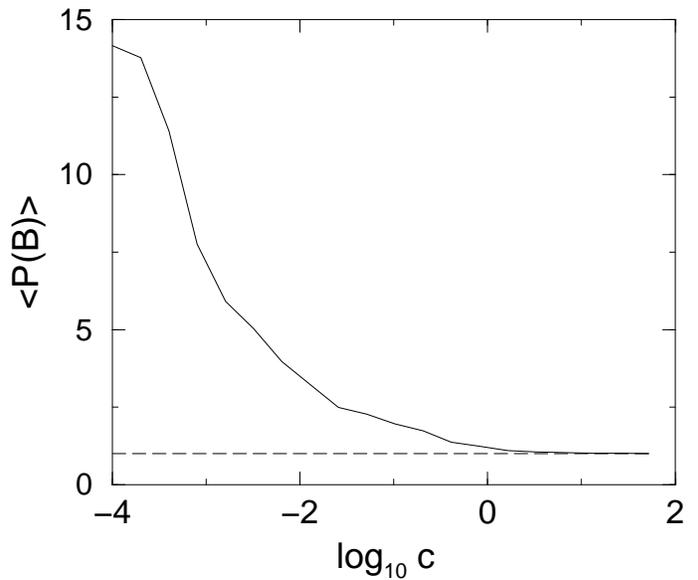}
\caption{The average value of the Peterlin function $P(B)$ as a function of
$c$ computed 
in the SabraP model. The dashed line corresponds to $P=1$.}
\label{P} 
\end{figure}

Viscoelastic flows are represented well by hydrodynamic equations in which
the effect of the 
polymer enters in the form of a ``conformation tensor" $\B R(\B r,t)$ which
stems from 
the ensemble average of the dyadic product of the end-to-end distance of the
polymer chains \cite{87BCAH,94BE}. A
successful model that had been employed frequently in numerical simulations
of turbulent 
channel flows is the FENE-P model. Flexibility and
finite extendability of the polymer chains are reflected by the relaxation
time $\tau$ and 
the Peterlin function $P(\B r,t)$ which appear in the equation of motion for
$\B R$: 
\begin{eqnarray} 
\frac{\partial R_{\alpha\beta}}{\partial t}+(\B u\cdot \B \nabla)
R_{\alpha\beta} 
&&=\frac{\partial u_\alpha}{\partial r_\gamma}R_{\gamma\beta}
+R_{\alpha\gamma}\frac{\partial u_\beta}{\partial r_\gamma}\nonumber\\
&&-\frac{1}{\tau}\left[ P(\B r,t) R_{\alpha\beta} -\rho_0^2
\delta_{\alpha\beta} \right]\label{EqR}\\
P(\B r,t)&&=(\rho_m^2-\rho_0^2)/(\rho_m^2 -R_{\gamma\gamma})
\end{eqnarray} 
In these equations $\rho^2_m$ and $\rho^2_0$ refer
to the maximal and the equilibrium values of the trace
$R_{\gamma\gamma}$. Since in most applications $\rho_m\gg \rho_0$ the
Peterlin function can also be
written approximately as $P(\B r,t)\approx (1/(1 -\alpha R_{\gamma\gamma})$
where $\alpha=\rho_m^{-2}$.
In its turn the conformation tensor appears in the equations for the fluid
velocity $\B u(\B r,t)$ as an
additional stress tensor:
\begin{eqnarray} 
&&\frac{\partial \B u}{\partial t}+(\B u\cdot \B \nabla) \B u=-\B \nabla p
+\nu_s \nabla^2 \B u +\B \nabla \cdot \B {\C T}+\B F\ , \label{Equ}\\
&&\B {\C T}(\B r,t) = \frac{\nu_p}{\tau}\left[\frac{P(\B r,t)}{\rho_0^2} \B
R(\B r,t) -\B 1 \right] \ . \label{T}
\end{eqnarray} 
Here $\nu_s$ is the viscosity of the neat fluid, $\B F$ is the forcing and
$\nu_p$ is a viscosity parameter which is related to the concentration of
the polymer, 
i.e. $\nu_p/\nu_s\sim\Phi$ where $\Phi$ is the volume fraction of the
polymer. 
We note however that the tensor field can be rescaled to get rid of the
parameter $\rho_m^2$ in the
Peterlin function, $\tilde R_{\alpha\beta}=\alpha R_{\alpha\beta}$ with the
only consequence 
of rescaling the parameter $\rho_0$ accordingly. 
Thus the actual value of the concentration 
is open to calibration against the experimental data.
These equations were simulated on the computer in a channel or pipe
geometry, reproducing faithfully the characteristics of drag reduction in
experiments. 
It should be pointed out however that even for present day computers
simulating these equations is quite tasking. It makes sense therefore to try
to 
model these equations further. For the purpose of studying drag reduction in
homogeneous systems
one can derive a shell model whose simplicity and transparency are assets
for analysis and 
simulations alike. 
In developing a simple model one is led by the following ideas. First, it
should 
be pointed out that all the nonlinear terms involving the tensor field
$\B R(\B r,t)$ can be reproduced by writing an equation of motion for a
vector field $\B B(\B r,t)$,
and interpreting $R_{\alpha\beta}$ as the dyadic product $B_\alpha B_\beta$.
The relaxation terms
with the Peterlin function are not automatically reproduced this way, and
one needs to add them
by hand. Second, we should keep in mind that the above equations exhibit a
generalized energy which
is the sum of the fluid
kinetic energy and the polymer free energy. Led by these consideration the
following shell model
was proposed in \cite{03BDGP,03BCHP}:
\begin{eqnarray} 
\frac{d u_n}{d t} &=& \frac{i}{3} \Phi_n (u,u) - \frac{i}{3}
\frac{\nu_p}{\tau} P(B)
\Phi_n (B,B) - \nu_s k^{2}_n u_n + F_n , \nonumber\\
\frac{d B_n}{d t} &=& \frac{i}{3} \Phi_n (u,B) - \frac{i}{3}
\Phi_n (B,u) - {1 \over \tau} P(B) B_n - \nu_B k_n^2 B_n,\nonumber\\
P(B) &=& {1 \over 1 - \sum_n B_n^* B_{n} } \ . \label{SP}
\end{eqnarray} 
In these equations $u_n$ and $B_n$ stand for the Fourier amplitudes $u(k_n)$
and $B(k_n)$ of the two
respective vector fields, but as usual in shell model we take
$n=0,1,2,\dots$ and the
wavevectors are limited to the set $k_n=2^n$. The nonlinear interaction
terms take 
the explicit form 
\begin{eqnarray} 
\Phi_n(u,B) &&= 3 \Big[k_n u_{n+1}^* B_{n+2} + b k_{n-1} u_{n-1}^*
B_{n+1} \nonumber\\&&+ (1+b) k_{n-2} u_{n-2}B_{n-1} \Big]\ ,
\end{eqnarray} 
with $b$ a parameter and
the obvious extension to $\Phi_n(u,u)$, $\Phi_n(B,u)$ and
$\Phi_n(B,B)$. 
In accordance with the generalized energy of the FENE-P model
\cite{87BCAH,94BE}, also our
shell model has the total energy
\begin{equation} 
E \equiv {1 \over 2} \sum_n |u_n|^2 - {1 \over 2} {\nu_p \over \tau}
\ln \left(1-\sum_n |B_n|^2\right)\ .
\end{equation} 
The second term in the generalized energy contributes to the dissipation a
positive definite 
term of the form $(\nu_p/\tau^2)P^2(B) \sum_n|B_n|^2$. With $\nu_p=0$ the
first of 
Eqs. (\ref{SP}) reduces to the well-studied Sabra model of Newtonian
turbulence. 
We therefore refer the model with $\nu_p \ne 0$ as the SabraP model.
As in the FENE-P case we
consider $c\equiv \nu_p/\nu_s$ to be proportional to the concentration of
polymers. 
In \cite{03BDGP} it was shown that this shell model exhibits drag reduction,
and the mechanism 
for the phenomenon was elucidated. Furthermore, it was shown in
\cite{03BCHP} that for large enough
concentration, the Peterlin function can be disregarded (i.e. $P \approx 1$)
and, consequently, the dynamics of the system
becomes concentration independent, i.e. we reach the MDR asymptote.
This behavior of the Peterlin function is shown in Fig. \ref{P}.
Following the above finding, we consider below the limiting case in which
the 
concentration is large enough for the Peterlin function to be close to
unity, $P\approx 1$.
Finally, all the numerical simulations reported in this paper have been
performed by using $b = -0.2$, $\nu_s = 10^{-7}$
and a constant energy input given by:
\begin{equation} 
F_n = \frac{10^{-3}}{u_n^*}
\end{equation} 
for $n=1,2$ and $F_n =0$ for $n > 2$.
\section{Scale dependent effective viscosity in homogeneous drag reduction}
\label{homo} 
\begin{figure} 
\centering 
\includegraphics[width=.5\textwidth]{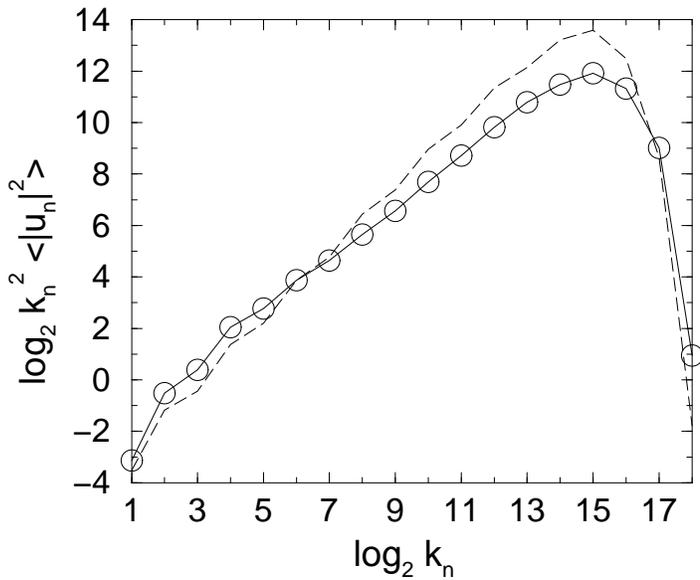}
\caption{The spectrum of the energy dissipation $k_n^2 e(k_n)$ for the Sabra
(solid line with symbol) and the SabraP models (dashed line). 
Both models show the same maxima for about $n\sim 15$ 
which corresponds to the peak in the energy dissipation.}
\label{fig3} 
\end{figure}

\begin{figure} 
\centering 
\includegraphics[width=.5\textwidth]{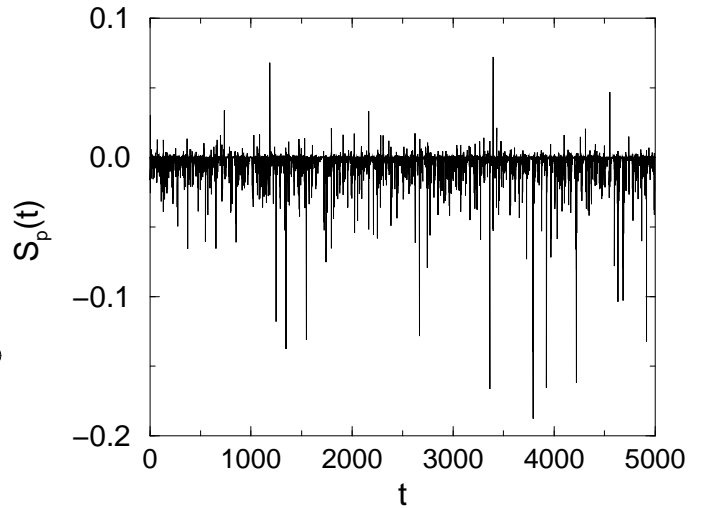}
\caption{Time behavior of $S_p$, as defined in Eq. (\ref{sp}), which
represents the whole energy exchange from the
$u_n$ field to $B_n$. Negative values of $S_p$ means that energy is taken
from $u_n$.} 
\label{fig2} 
\end{figure}

\begin{figure} 
\centering 
\includegraphics[width=.5\textwidth]{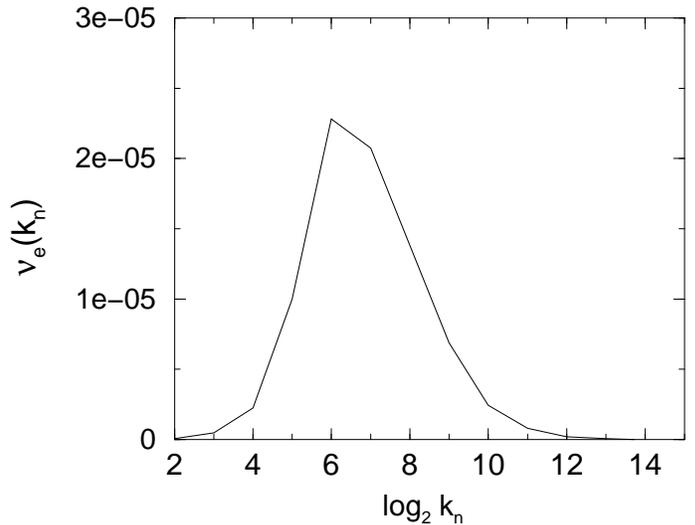}
\caption{The values of the eddy viscosity $\nu_e(k_n)$ defined in 
Eq. (\ref{nue}) for $P(B)=1$.
Note that this quantity rises rapidly in the vicinity of the Lumley scale.}
\label{fig4} 
\end{figure}

\begin{figure} 
\centering 
\includegraphics[width=.5\textwidth]{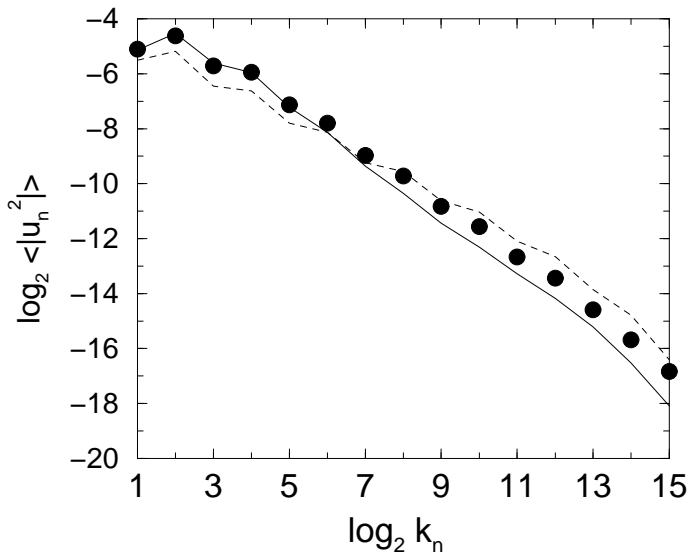}
\caption{The energy spectrum
of the SabraP model (solid line) as compared with the energy spectrum
of the Sabra model with the effective viscosity and
$\alpha = 0.3$ (symbols) and the Sabra model without effective viscosity
(dashed line).} 
\label{fig5} 
\end{figure}

\begin{figure} 
\centering 
\includegraphics[width=.5\textwidth]{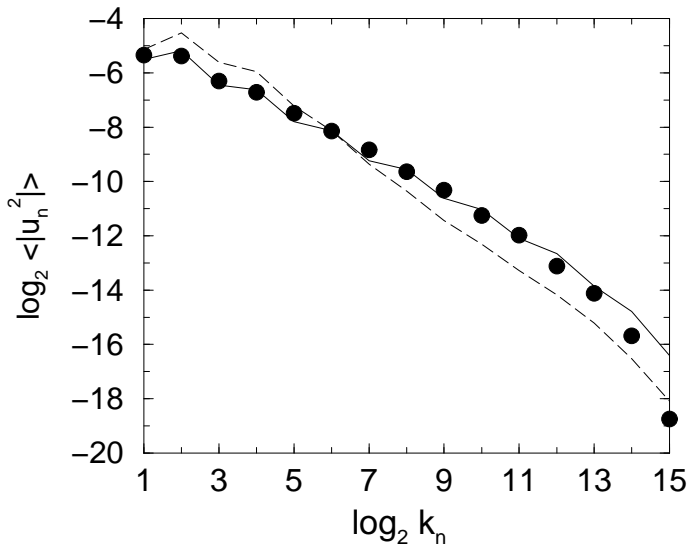}
\caption{Energy spectrum of the Sabra model for $\nu_s$ (line), the Sabra
model with increased viscosity $\nu^*$ (symbols) and for the Sabra
model with an effective
viscosity $0.3\nu_e(k_n)$ (dashed line).}
\label{fig6} 
\end{figure}

\begin{figure} 
\centering 
\includegraphics[width=.5\textwidth]{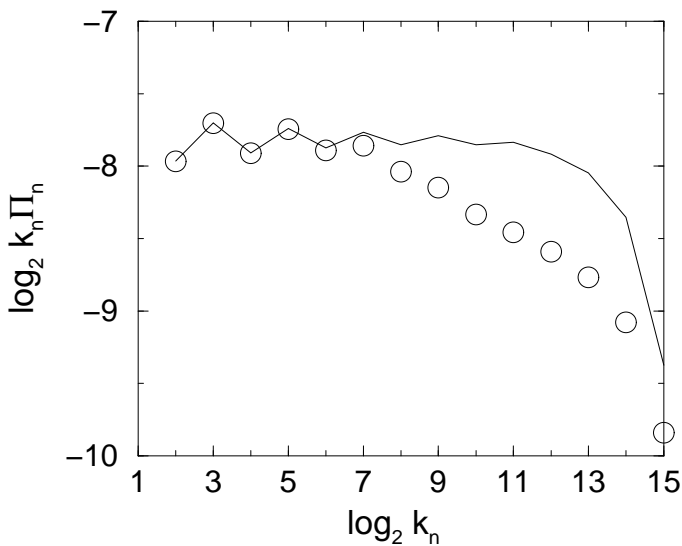}
\caption{Energy flux computed for the SabraP model (solid line) and the
Sabra model with effective viscosity (symbols).} 
\label{fig7} 
\end{figure}

Drag reduction in homogeneous turbulence is exhibited by a relative increase
in the rms 
fluctuations of the energy at large scales. We thus focus naturally on the
energy spectrum 
$e(k_n) \equiv \langle u_nu_n^* \rangle$. In the context of the shell model
the phenomenon 
is demonstrated in Fig. \ref{spec} where $e(k_n)$ is shown for the given
values of the parameters.
The spectra for the pure Sabra model (line with symbols) and the coupled
model (line) are compared
for the same amount of power input per unit time. The discussion of the
spectra revolves around the
typical Lumley scale $k_c$ which is determined by the condition
\begin{equation} 
e^{1/2}(k_c) k_c\approx \tau^{-1} \la P(B)\ra\ . \label{defkc}
\end{equation} 
For $k_n\gg k_c$ the decay time $\tau$ becomes irrelevant for the dynamics
of $B_n$. The nonlinear
interaction between $u_n$ and $B_n$ at these scales results in both of them
having the same spectral
exponent which is also the same as that of the pure Sabra model. The
amplitude of the $u_n$ spectrum
is however smaller than that of the pure Sabra in the coupled case, since
the $B_n$ field adds
to the dissipation. On the other hand, for $k_n\ll k_c$ the $B_n$ field is
exponentially suppressed by
its decay due to $\tau$, and the spectral exponent of $u_n$ is again as in
the pure Sabra. Drag reduction
comes about due to the interactions at length scales of the order of $k_c$
which force a strong tilt
in the $u_n$ spectrum there, causing it to cross the pure Sabra spectrum,
leading to an increase in the
amplitude of the energy containing scales. This is why the kinetic energy is
increasing for the same
amount of power input, and hence drag reduction. In Fig. \ref{fig3} we
show the spectrum of energy dissipation $k_n^2 e(k_n)$. This figure
indicates that as far as the dissipative scale
is concerned, it is not changed much by the coupling of the velocity field
to the polymer field;
both models show a maximum at $n\sim 14$ 
which is the dissipative scale. We
now address the question
how to recapture the same phenomenon in a model involving the velocity field
alone but with a scale-dependent effective viscosity.
We first reiterate that the field $u_n$ loses energy in favor of the field
$B_n$. 
Using Eq. (\ref{SP}) we can measure the energy transfer from $u_n$ to $B_n$
using the quantity:
\begin{equation} 
S_p\equiv \sum_n s_p(k_n)\equiv \frac{\nu_p P(B)}{3\tau}{\rm Re} \{i\Sigma_n
u^*_n \Phi_n(B,B)\} \ .
\label{sp} 
\end{equation} 
This function measures the exchange between the kinetic energy $\Sigma_n
u^*_nu_n$ and the 
``polymer'' or ``elastic''
energy $\Sigma_n B_n^*B_n$.
In Fig. \ref{fig2} we show a snapshot of the dependence of the function
$S_p$ on time. 
The point to notice is that $S_p$
is negative for most of the time. The $B_n$ field drains
energy from the velocity field, and we therefore can hope to be able to
capture its role by an effective viscosity. Note however that the dynamics
of $S_p$ 
is strongly intermittent; this feature is common to the shell model and the
full FENE-P model as observed in the DNS of the latter. We cannot hope to
capture {\em all} the
temporal complexity with the notion of effective viscosity, since the latter
is an 
average notion. Nevertheless the essential features 
will be shown to be reproduced.
We will try to capture the effect of $S_p$ in terms of an effective
viscosity as follows:
using $\la .. \ra$ for the (time) average, we introduce the scale dependent
effective viscosity
$\nu_e(k_n)$ as: 
\begin{equation} 
\nu_e(k_n) = \frac{\la s_p(k_n)\ra }{k_n^2 e(k_n)} \ .
\label{nue} 
\end{equation} 
The quantity $\nu_e(k_n)$ is shown in Fig. \ref{fig4}; its
maximum is reached at $n\sim 6-7$, 
a wavenumber which is not yet in the
dissipative range. 
It is important to stress that $\nu_e(k_n)$ is obtained by averaging over a
complex and intermittent dynamical
behavior of the viscoelastic shell model. It is therefore not obvious that
the main characteristics of
drag reduction can be obtained by simply replacing the viscoelastic terms
$\Phi_n(B,B)$ by a scale
dependent effective viscosity. We demonstrate that this is possible by
using now the Sabra model with an extra viscous
term given by $\nu_e(k_n)k_n^2 u_n$. The new viscous term replaces, on the
average, the effect of viscoelastic terms
proportional to $\Phi_n(B,B)$. The equations of motion read:
\begin{equation} 
\frac{d u_n}{d t} = \frac{i}{3} \Phi_n (u,u)
-\nu_e(k_n) k_n^2 u_n - \nu_s k^{2}_n u_n + F_n \ .
\label{sabrae} 
\end{equation} 
We do not expect that $\nu_e(k_n)$ in the {\em dynamics} of the Sabra model,
as proposed in Eq. (\ref{sabrae}), will be exactly the object measured on the
average defined in Eq.~(\ref{nue}).
We clearly must keep the functional dependence of $\nu_e$ on $k_n$, but we
can allow a factor of proportionally
that will take care of the difference between the dynamical intermittent
behavior and the average
behavior. We will therefore use the form $\alpha
\nu_e(k_n)$, where $\alpha$ is a constant that can be optimized to achieve a
close correspondence between
the two-field model and the effective one-field model. For $\alpha=0$ we
recapture 
the original Sabra model without effective viscosity.
We simulated the Sabra model with
the effective viscosity [Eq.~(\ref{sabrae})] 
for different values of $\alpha$ in the range $(0,1)$.
Drag reduction was found in all cases.
For $\alpha = 0.3$ the energy spectrum turns out to be very close to the
original SabraP model
{\em with the viscoelastic terms}. In Fig. \ref{fig5} we show the energy
spectrum of the SabraP model and the energy
spectrum of the 
Sabra model with effective viscosity for $\alpha =0.3$.
In order to check that the result shown in Fig. \ref{fig5} are due to a
{\em scale dependent} dissipation, we have
defined a {\em scale independent viscosity} $\nu^*$ as:
\begin{equation} 
\nu^* = \frac{\la S_p\ra}{\Sigma_n k_n^2 e(k_n)}
\label{nustar} 
\end{equation} 
The definition of $\nu^*$ is similar to that given in Eq. (\ref{nue}), i.e.
$\nu^*$ is defined such that, by adding a
viscous term $\nu^* k_n^2 u_n$ to the Sabra model, the system {\em on the
average} is losing the same amount of energy
as in the case of viscoelastic flows.
It turns out that in our case $\nu^* \sim 2.5\times 10^{-7}$. By using this
value for $\nu^*$ we
have numerically integrated the Sabra model by adding a new viscosity equal
to $\nu^*$, namely:
\begin{equation} 
\frac{d u_n}{d t} = \frac{i}{3} \Phi_n (u,u)
-\nu^* k_n^2 u_n - \nu_s k^{2}_n u_n + F_n
\label{sabrastar} 
\end{equation} 
The corresponding energy spectrum is shown in Fig. \ref{fig6}
together with the energy spectrum for the Sabra model (viscosity $\nu_s$)
and the Sabra model with the effective viscosity
$0.3\nu_e(k_n)$. As one can clearly see, an increase of the dissipation for
all scales does not result in a drag reduction.
Finally, we have computed the energy flux of the Sabra model with effective
dissipation and compared it against the energy flux of
the SabraP model. This comparison is exhibited in Fig. \ref{fig7} where the
solid line  corresponds to the SabraP model and the
symbols correspond to
the Sabra model with effective dissipation $0.3\nu_e(k_n)$. The two energy
fluxes are equal in the inertial range up to
wavenumber $n \sim 7$.

The results illustrated so far support the conclusion that a scale
dependent effective viscosity is able to reproduce most
of the dynamics of viscoelastic terms and, in particular, the phenomenon of
drag reduction. Let us remark once more, that
it is the {\em scale dependence} of the effective 
viscosity which is able to properly
reproduce the drag reduction. It is worthwhile to explain the mechanism
of the action of the scale dependent viscosity, to understand its
similarity to the action of the polymers.  For fixed energy input,
as in our 
case, drag reduction is shown as an increase of the rms fluctuations at
scales larger than the Lumley scale.
The scale dependent effective viscosity increases the viscous terms $k_n^2
u_n$ in a particular range of scales, say
for $n_c < n < n_2$, where $n_c =\log_2 (k_c)$.  The energy flux $\Pi_n$
in the system is given by the third order correlation function

$\Pi_n \sim \la u^*_{n-1}u^*_nu_{n+1}\ra$.
As shown in the Fig. \ref{fig7}, we can safely assert that the energy flux
does not 
change for 
$n<n_c$. The increase of viscosity at $n=n_c$ produces a decrease of the
energy at scale $n_c$. Thus, we expect
$u_{n_c}$ to decrease with respect to the value observed in the Newtonian
case. Since
$\Pi_n$ is not affected by the increase of the viscosity at $n=n_c$, we must
conclude that the quantity $u_{n-1}u_n$ should
increase while $u_{n_c}$ decreases. 
This is the origin of the tilt in the An increase of $u_n$
spectrum in the vicinity of $n_c$.  From a physical point of view, this
picture is not 
different 
from the one discussed in \cite{03BDGP} where a similar
explanation for the drag reduction was invoked.  Note that all that we need
for the phenomenon to occur is that the increase
in viscosity should start at the right scale. This scale is equivalent of
the 
Lumley scale whose role in the viscoelastic case had been already
emphasized. 

\begin{figure} 
\centering 
\includegraphics[width=.5\textwidth]{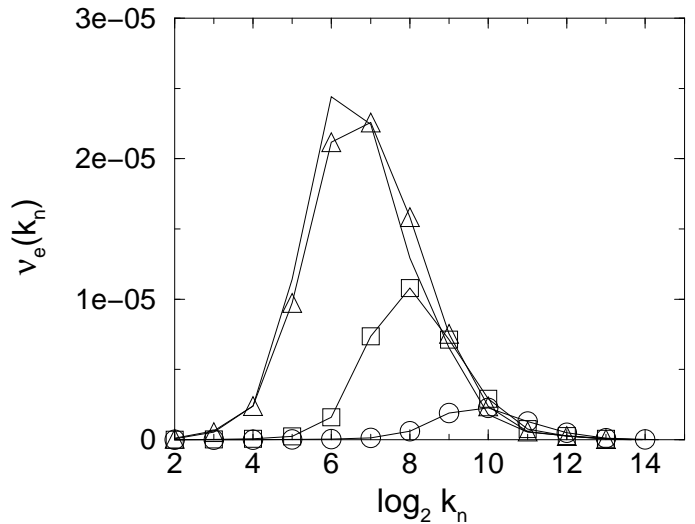}
\caption{Effective viscosity for varying the concentration: $c=10^{-2}$
(circles), 
$c=10^{-1}$ (squares), $c=10$ (triangles) and $c=100$ (line).}
\label{fig8} 
\end{figure}

Finally, we discuss the effect of changing the concentration on the
effective viscosity.
When $\la P(B)\ra >1$ the effective viscosity depends on the Peterlin
function, 
which in turn 
depends on the concentration $c$ and on the relaxation time $\tau$, cf. Eq.
(\ref{nue}). 
Figure \ref{fig8} displays the effective viscosity as a function of $k_n$ for
four values 
of the concentration, $c=10^{-2}$, $10^{-1}$, 10 and 100. As the
concentration decreases,
the effective viscosity decreases, and its peak migrates to higher values of
$k_n$. This 
migration is simply due to the change in the Lumley scale, cf.. Eq.
(\ref{defkc}). The 
decrease in the effective viscosity is due to the increase in $\la P(B)\ra$
shown in Fig. \ref{P}. Needless
to say, these changes in the effective viscosity decrease the effect of
drag reduction, as seen in experiments and simulations: only large
concentrations agree with the MDR asymptote.
\section{The limit of large system size}
\label{limit} 

In this section we want to discuss the limit $k_0 \rightarrow 0$ while
keeping fixed the
scale and the shape of the effective viscosity. In other words, we study
$k_0 \rightarrow 0$ for fixed value of the Lumley scale $k_c$. Note that we
take $k_c$ much
smaller than the dissipative scale and we keep constant the rate of energy
input
$\epsilon$. 

\begin{figure} 
\centering 
\includegraphics[width=.5\textwidth]{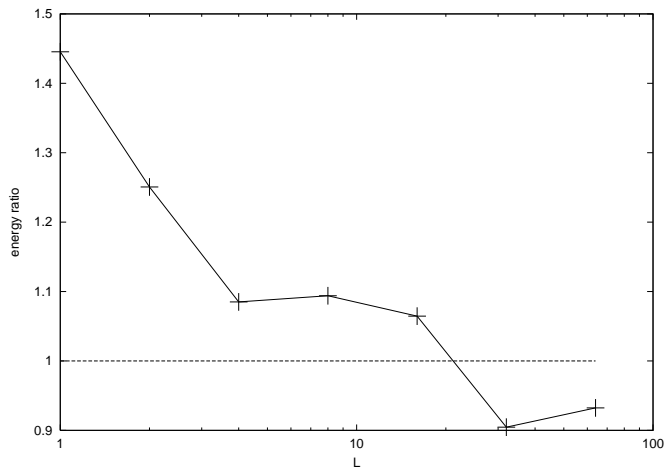}
\caption{Ratio of the kinetic energy for the Sabra model with scale dependent 
viscosity and the kinetic energy of ths Sabra model with fixed kinematic viscosity, 
for different values of $L\equiv k_0^{-1}$.
Note that for drag reduction to take place the ratio must be larger than $1$. 
The position the maximum in the scale dependent
viscosity is kept fixed while $L \rightarrow \infty$. }
\label{fig13} 
\end{figure}

The discussion simplifies by considering the other typical scale in our
system, which is the
Taylor microscale $\lambda_T$,
\begin{equation}
\lambda_T \equiv \sqrt{\frac{\sum_n\langle |u_n|^2\rangle}{\sum k_n^2
|u_n|^2}} \ . \label{deflamT}
\end{equation}
In \cite{03BDGP} it was shown that the conditions are optimal for drag
reduction in our shell model when a dimensionless
parameter $\mu\equiv \lambda_Tk_c$, is of the order of unity. On the
other hand drag reduction
is lost when $\mu\gg 1$ or $\mu\ll 1$. Obviously, when $k_0\to 0$ the
overall kinetic energy
increases as $k_0^{-2/3}$ while the denominator in Eq. (\ref{deflamT}) 
remains
unchanged, being dominated
by the viscous scale. Thus $k_0\to 0$ leads to $\lambda_T\to \infty$, and we
expect to lose drag reduction
in that limit (for a fixed value of $k_c$). This
conclusion is supported by the results shown in Fig.~\ref{fig13}, where we
plot the 
ratio between the kinetic energy with the effective viscosity and the
Newtonian kinetic energy
for $L \equiv k_0^{-1} \rightarrow \infty$. The case $L=1$ corresponds to
the previous sections.
Note that for $L$ large enough, the system exhibits drag
enhancement. Physically, for very large values of $k_c/k_0$ the effective
dissipation is just increasing the overall viscosity in the system and,
therefore, no drag reduction can be
observed. 
For drag reduction to occur we must have the
Lumley scale close to energy containing scales. Note, however, that
``close'' 
in our case means $k_c \sim 50 - 100 $ larger than the integral scale $k_0$.

It is interesting to compare our findings, which pertain to homogeneous
systems, to drag reduction in turbulent boundary
layers. The elastic layer in such flows (between the viscous layer and the
Newtonian plug) has the peculiar distinction
that $y$, the distance from the wall, becomes the only important scale in
the problem. It is both the energy containing
scale and the Lumley scale at the same time. The former is clear; at
distance $y$ from the wall the most
energetic eddies are of size $y$. The latter needs a bit of theory, and this
is provided in \cite{03LPPT}. The upshot
of the analysis there is that in the elastic layer the kinetic energy $K(y)$
scales like $K(y)\sim y^2/\tau^2$.
Thus the Lumley scale is also $y$. Accordingly, the phenomenon of drag
reduction is totally indifferent to
the physical size of the channel (or pipe). As long as the conditions for
drag reduction hold at distance $y$
from the wall, drag reduction will occur and will have a persistent effect
on the mean flow independently of
the outer scale. 
Eventually, when $y$ is large enough, $K(y)$ may stop growing like $y^2$,
the Lumley scale decreases, and we observe
cross over back to the
Newtonian log layer, albeit shifted to a larger value of a mean velocity
profile. 

In summary, drag reduction phenomena in homogeneous and wall bounded flows
have a lot in common even
though the effect disappears in the former when the system size goes
to infinity. The essential
physics is the proximity
of the Lumley scale to the energy containing scales, which allows an
effective interaction between
the polymer dynamics and the hydrodynamic modes.

\section{Conclusions}
The work presented in this paper supports two conclusions. First, we
demonstrated 
that drag reduction by polymers can be represented in terms of an effective
scale dependent viscosity.  One
can use a theory in which two fields are explicitly presented, i.e. the
velocity field and the polymer field. Then the viscosity remains Newtonian,
and the 
polymer conformation tensor acts as the additional sink of energy at the
intermediate scales which are larger than the viscous scales but smaller
than the Lumley 
scale. We can construct however effective models in which only the velocity
field 
is present, and replace the polymer field by an effective viscosity. This
effective viscosity will be different from the Newtonian one at the crucial
scales at 
which the polymers are active, i.e. scales larger than the dissipative
scales but 
smaller than the Lumley scale. With a properly chosen effective viscosity we
can 
reproduce the results of the two-field theory qualitatively and even
semi-quantitatively.
Having done so, we reach a unified discussion of drag reduction by polymers
in homogeneous 
and wall bounded flows. It is worth pointing out however that the unified
discussion is 
deeper than the device of unified viscosity. Superficially drag reduction in
homogeneous 
and wall bounded turbulence appear very different. In the former there is no
mean flow and 
drag reduction appears as an increase of the rms fluctuations of the large
scales. In the 
latter drag reduction means the increase of the mean flow velocity.
Nevertheless in essence
the phenomenon of drag reduction in homogeneous and wall bounded flows is
basically the same:
the polymers act to reduce the gradients at the intermediate scales. They
partly laminarize 
the flow at the intermediate scales, and this allows the largest scales
to attain higher rms fluctuation levels (in homogeneous flows) or higher
mean velocity 
(in wall bounded flows). To understand this further recall that for laminar
flows the 
drag is a strongly decaying function of Re. Once turbulence sets in, the
dramatic increase 
in eddy viscosity contributes to a drag which is much larger than the one
that would obtain 
in a hypothetical laminar flow with the same value of Re. The addition of
polymers allows one
to bring the drag closer to the hypothetical laminar low value, and this is
done by reducing the turbulence level at intermediate scales. Whether one prefers to
describe 
the quantitative aspects of this phenomenon using explicitly the polymer
field or by 
employing an effective viscosity depends to a large extent on one's goals.
We expect that the concept of effective viscosity will be found equally
useful in discussing
drag reduction in other situations, for example when microbubbles are used
instead of polymers.
The quantitative aspects of such a description need however to be worked out
case by case, and 
this is our program for the near future.
\acknowledgments 
\vskip 0.5 cm 
This work was supported in part by the European Commission
under a TMR grant, the US-Israel Binational Science Foundation, and the
Minerva Foundation, Munich, Germany. ESCC was supported by a grant of the Research Grant
Council of Hong Kong
(Ref. No. CUHK 4046/02P).

\end{document}